# Latency Optimized Asynchronous Early Output Ripple Carry Adder based on Delay-Insensitive Dual-Rail Data Encoding

P. Balasubramanian, and K. Prasad

*Abstract*—Asynchronous circuits employing delay-insensitive codes for data representation i.e. encoding and following a 4-phase return-to-zero protocol for handshaking are generally robust. Depending upon whether a single delay-insensitive code or multiple delay-insensitive code(s) are used for data encoding, the encoding scheme is called homogeneous or heterogeneous delay-insensitive data encoding. This article proposes a new latency optimized early output asynchronous ripple carry adder (RCA) that utilizes single-bit asynchronous full adders (SAFAs) and dual-bit asynchronous full adders (DAFAs) which incorporate redundant logic and are based on the delay-insensitive dual-rail code i.e. homogeneous data encoding, and follow a 4-phase return-to-zero handshaking. Amongst various RCA, carry lookahead adder (CLA), and carry select adder (CSLA) designs, which are based on homogeneous or heterogeneous delay-insensitive data encodings which correspond to the weak-indication or the early output timing model, the proposed early output asynchronous RCA that incorporates SAFAs and DAFAs with redundant logic is found to result in reduced latency for a dual-operand addition operation. In particular, for a 32-bit asynchronous RCA, utilizing 15 stages of DAFAs and 2 stages of SAFAs leads to reduced latency. The theoretical worst-case latencies of the different asynchronous adders were calculated by taking into account the typical gate delays of a 32/28nm CMOS digital cell library, and a comparison is made with their practical worst-case latencies estimated. The theoretical and practical worst-case latencies show a close correlation.

The proposed early output 32-bit asynchronous RCA, which contains 2 stages of SAFAs in the least significant positions and 15 stages of DAFAs in the more significant positions, reports the following optimizations in latency over its architectural counterparts for a similar adder size: i) 35.3% reduction in latency over a weak-indication section-carry based CLA (SCBCLA), ii) 30.5% reduction in latency over a weak-indication hybrid SCBCLA-RCA, iii) 20.2% reduction in latency over an early output recursive CLA (RCLA), iv) 18.7% reduction in latency over an early output hybrid RCLA-RCA, and v) a 13% reduction in latency over an early output CSLA that features an optimum 8-8-8-8 uniform input partition.

P. Balasubramanian was with the School of Computer Science and Engineering, Nanyang Technological University, Singapore. He is now with the School of Electrical and Electronic Engineering, Nanyang Technological University, Singapore 639798 (e-mail: balasubramanian@ntu.edu.sg)

K. Prasad is with the Department of Electrical and Electronic Engineering, Auckland University of Technology, Auckland 1142, New Zealand (e-mail: krishnamachar.prasad@aut.ac.nz).

## I. INTRODUCTION

THE binary full adder is a fundamental arithmetic component used in microprocessor, microcontroller, and digital signal processing units. The full adder is used to add two binary inputs along with a carry input from a preceding stage and produces two binary outputs, i.e. the sum and the carry output. The full adder can be realized in synchronous [1] – [4] or asynchronous design style [5] – [15]. As an alternative to the conventional single-bit full adder, the dual-bit full adder was proposed in [16] – [18] based on the synchronous and asynchronous design paradigms. The dual-bit full adder adds two augend and addend binary inputs along with a carry input and produces two sum outputs and also a carry output. It was shown in [16] – [18] that irrespective of whether the design style adopted is synchronous or asynchronous, the dual-bit full adder when cascaded to form a RCA would substantially reduce the latency, i.e. the critical path delay of a RCA that is constructed using only single-bit full adders.

In this work, we present the novel design of an early output asynchronous RCA that is robust and incorporates DAFAs and SAFAs which are based on the delay-insensitive dual-rail data encoding scheme (i.e. homogeneous data encoding) and corresponds to the 4-phase return-to-zero handshaking. The proposed asynchronous RCA when used to perform 32-bit addition results in reduced latency compared to its counterpart RCAs comprising single-bit and/or dual-bit full adders which are based on homogeneous or heterogeneous delay-insensitive data encoding. Also, the proposed 32-bit asynchronous RCA reports the optimized latency compared to its architectural counterparts such as CLA and CSLA. This inference is based on the simulation results obtained by using a 32/28nm CMOS process.

The remainder of this research article is organized into five sections. A relevant background about robust asynchronous design that is based on delay-insensitive data encoding (homogeneous or heterogeneous), and which adheres to the 4-phase (return-to-zero) handshake protocol is provided in Section II. The proposed asynchronous RCA design is discussed in Section III. Next, the simulation results corresponding to various 32-bit asynchronous RCAs, CLAs, and a CSLA are presented and compared in Section IV. The mathematical estimation of the theoretical latencies of various 32-bit adders and a comparison with their practical latency values are also given in this section. Finally, the conclusions are drawn in Section V.

## II. ROBUST ASYNCHRONOUS CIRCUITS – DESIGN FUNDAMENTALS

### A. Asynchronous Circuit Stage Operation

An asynchronous function block is the equivalent of the synchronous combinational logic [19]. When an asynchronous function block is constructed using delay-insensitive codes [20] and utilizes a 4-phase (return-to-zero) handshaking, it is generally robust provided it is free of gate and wire orphans [21] – [23]. Orphans are unacknowledged signal transitions which may occur on gate outputs (gate orphans) or in wires (wire orphans). Wire orphans are eliminated by imposing the isochronicity assumption [24], which forms the weakest compromise to delay-insensitivity. An isochronic fork implies that a signal transition on a wire junction i.e. a node is concurrently transmitted across all the wire branches. Gate orphans may however become problematic and so their possibility of occurrence should be eliminated to guarantee that an asynchronous circuit based on delay-insensitive data encoding and 4-phase handshaking would be robust.

The dual-rail code, also called 1-of-2 code, is the simplest member of the family of delay-insensitive $m$-of-$n$ data codes [20]. Among the family of $m$-of-$n$ codes, the 1-of-$n$ codes represent a subset and are called one-hot codes. In a 1-of-$n$ code, only 1 out of $n$ wires is asserted high, i.e. binary 1 to represent a binary data. In fact, the 1-of-$n$ coding scheme is said to be unordered [25] since none of the code words forms a subset of any other code word. Further, the 1-of-$n$ coding scheme is said to be complete [26] if all the $n$ unique code words as per the definition are utilized to encode the specified binary data. Table 1 shows an example binary data represented using the 1-of-2 and 1-of-4 data encoding schemes.

TABLE 1. 2-BIT BINARY DATA REPRESENTED USING 1-OF-2 AND 1-OF-4 DATA ENCODING SCHEMES

| Binary data | | 1-of-2 encoded data | | 1-of-4 encoded data | | | |
|---|---|---|---|---|---|---|---|
| P | Q | (P1,P0) | (Q1,Q0) | F0 | F1 | F2 | F3 |
| 0 | 0 | (0,1) | (0,1) | 1 | 0 | 0 | 0 |
| 0 | 1 | (0,1) | (1,0) | 0 | 1 | 0 | 0 |
| 1 | 0 | (1,0) | (0,1) | 0 | 0 | 1 | 0 |
| 1 | 1 | (1,0) | (1,0) | 0 | 0 | 0 | 1 |

As per the 1-of-2 code, a single-rail binary input, say X, is encoded using two wires, say X1 and X0, where the data X = 1 is represented by X1 = 1 and X0 = 0, and the data X = 0 is represented by X1 = 0 and X0 = 1. X1 and X0 cannot assume 1 simultaneously as it is illegal and invalid because the coding scheme will not remain unordered. However, X1 and X0 can assume 0 simultaneously and this is referred to as the spacer. Hence as per the 1-of-2 code, a valid data is specified by either X1 or X0 assuming binary 0 and the other assuming binary 1, and the condition of both X1 and X0 assuming binary 0 is labelled the spacer or null i.e. no data. On the other hand, the 1-of-4 code is used to represent two bits of binary information at a time. Referring to Table 1, it can be seen that the two binary inputs specified by P and Q are encoded into F0, F1, F2 and F3 as per the 1-of-4 code for an example. When just one delay-insensitive code (say, 1-of-2 code) is alone used to encode the given binary data, it is called homogeneous delay-insensitive data encoding, and when more than one delay-insensitive code (for example, 1-of-2 and 1-of-4 codes) is used to encode the specified binary data, it is called heterogeneous delay-insensitive data encoding.

A typical asynchronous circuit stage that employs delay-insensitive codes for data encoding and processing, and the 4-phase return-to-zero handshake protocol for data communication is shown in Fig 1. As the name suggests, the 4-phase return-to-zero handshake protocol consists of 4 phases. This will be explained with reference to Fig 1 based on the assumption that the dual-rail code is used for data representation. Nevertheless, the explanation would be applicable for data represented using any delay-insensitive 1-of-$n$ code.

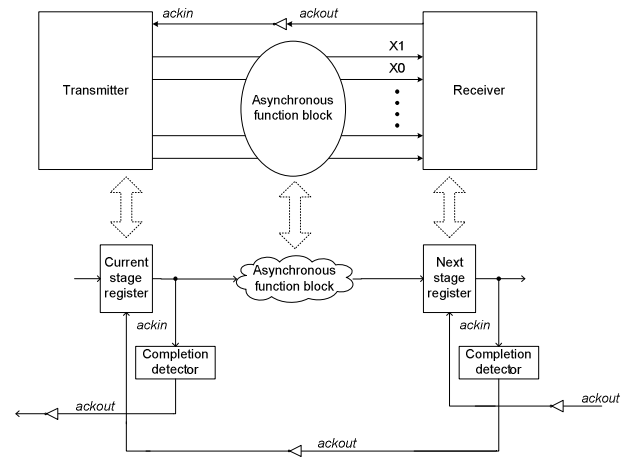

Fig 1 A robust asynchronous circuit stage operation that is correlated with the transmitter-receiver analogy for illustration

In the first phase, the dual-rail data bus shown in Fig 1 is in the spacer state and *ackin* is high i.e. binary 1. The transmitter now transmits a code word i.e. valid data and this results in upgoing signal transitions on any one of the corresponding dual rails of the entire dual-rail data bus. In the second phase, the receiver receives the code word sent, and it drives *ackout* high. In the next phase viz. the third phase, the transmitter waits for *ackin* to go low i.e. binary 0 and then resets the entire dual-rail data bus to the spacer state. Subsequently in the fourth phase, after an unbounded but a finite and positive time duration, the receiver drives *ackout* low i.e. *ackin* becomes high. One data transaction is now said to be completed, and the asynchronous circuit stage is ready to process the next data transaction.

The completion detector block [19], which is highlighted in Fig 1, ensures the complete arrival of all the primary inputs to an asynchronous circuit whether they are valid data or spacer.

It consists of an array of 2-input OR gates in the first logic level with each 2-input OR gate used to combine the respective dual-rails of an encoded primary input. The outputs of all the 2-input OR gates are synchronized using a C-element[1] tree, whose granularity depends on the composition of the digital cell library used for physical implementation.

*B. Asynchronous Function Block – Types*

Asynchronous function blocks are classified as strongly indicating, weakly indicating, and early output types. Indication implies acknowledging the arrival of the primary inputs to an asynchronous circuit through corresponding monotonic transitions on the intermediate and primary outputs. The transitions are expected to monotonically increase or decrease uniformly throughout the circuit [27]. The general input-output timing characteristics of strong-indication, weak-indication, and early output type asynchronous function blocks are portrayed through Fig 2.

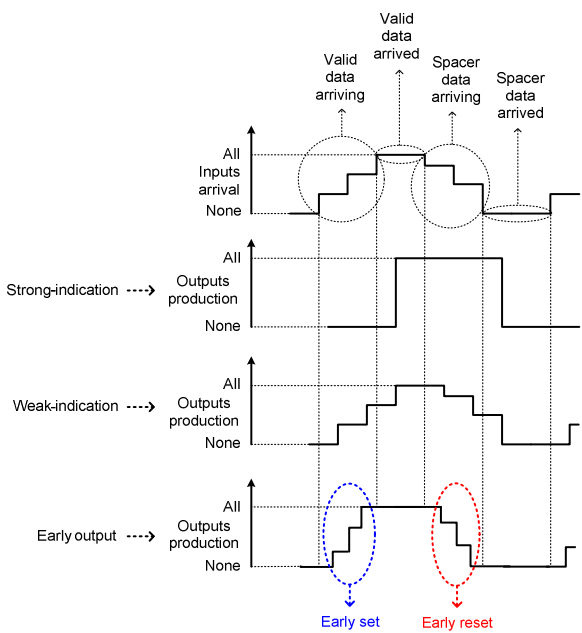

Fig 2 Depicting inputs-outputs timing correlation of strong-indication, weak-indication, and early output type asynchronous function blocks

A strong-indication function block [5] [28] starts data processing only after receiving all the primary inputs, and the required outputs are then produced.

A weak-indication function block [5] [29] is able to commence data processing after receiving just a subset of the primary inputs and can produce some of the primary outputs. However, only after receiving the last primary input, the last corresponding primary output is produced by the weak-indication function block. With respect to weak-indication, the mechanism may be either local or global [30]: local, if the asynchronous function block is internally indicating, and global, if the asynchronous circuit stage provides indication on the whole. It was shown in [31] that local weak-indication is preferable compared to global weak-indication for robust asynchronous circuit designs.

An early output function block [32] [33] is the most relaxed compared to the strong-indication and weak-indication function blocks as it can start data processing after receiving just a subset of the primary inputs and can produce all the primary outputs without having to wait for the arrival of all the primary inputs. In this context, the early output asynchronous function block could exhibit either early set or early reset behavior as shown in Fig 2. Early set implies that upon receiving a subset of the valid primary inputs, the early output function block produces all the valid primary outputs. The early set property is highlighted through the blue oval in dotted lines in Fig 2. On the other hand, the early reset behavior implies that upon receiving a subset of the spacer primary inputs, the early output function block processes them and drives all the primary outputs to the spacer state. The early reset property is shown by the red oval in dotted lines in Fig 2.

### III. ASYNCHRONOUS EARLY OUTPUT RCA COMPRISING SAFAs AND DAFAs

An asynchronous RCA architecture is proposed here which makes use of SAFAs and DAFAs instead of utilizing only SAFAs. Due to the usage of DAFAs, the number of carry propagation stages within the RCA is greatly reduced and so the data propagation delay i.e. the (forward) latency would be minimized. An example 32-bit asynchronous RCA that makes optimum use of SAFAs for the least significant positions and DAFAs for the more significant positions is shown in Fig 3.

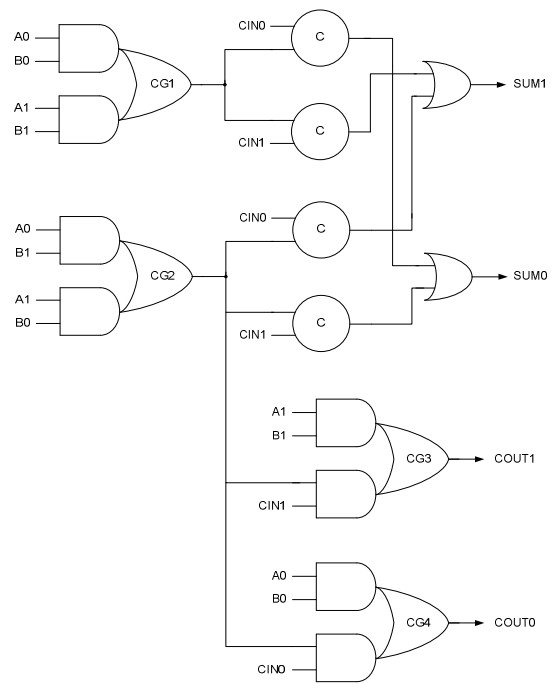

Fig 4 Early output SAFA [11] with implicit logic redundancy

---
[1] The C-element outputs binary 1 or 0 if all its inputs are binary 1 or 0 respectively. However, if its inputs are different, the C-element retains its existing steady-state output. The 2-input C-element is portrayed by a circle with the marking 'C' in the figures.

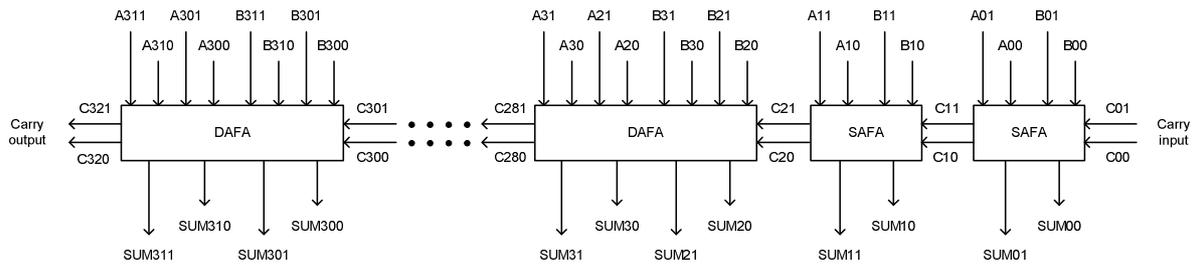

Fig 3 32-bit asynchronous RCA composed using SAFAs in the least significant positions and DAFAs in the more significant positions. SAFAs in the least significant adder positions help to reduce the significant latency that a least significant DAFA is likely to encounter. For example, the forward latency of two least significant SAFAs (Fig 4) equals the propagation delays of 3 AO22 gates, whereas the forward latency of a least significant DAFA (Fig 7) equals the sum of propagation delays of AND4, OR2, and 3 AO21 gates. In general, as per the proposed RCA structure, the DAFAs and SAFAs may correspond to strong-indication or weak-indication or early output timing models

$$SUM1 = A0B0CIN1 + A0B1CIN0 + A1B0CIN0 + A1B1CIN1 \quad (1)$$

$$SUM0 = A0B0CIN0 + A0B1CIN1 + A1B0CIN1 + A1B1CIN0 \quad (2)$$

$$COUT1 = A0B1CIN1 + A1B0CIN1 + A1B1CIN0 + A1B1CIN1 \quad (3)$$

$$COUT0 = A0B0CIN0 + A0B0CIN1 + A0B1CIN0 + A1B0CIN0 \quad (4)$$

Fig 5 Logic equations corresponding to the dual-rail encoded SAFA. The equations are factorized [42] according to the safe quasi-delay-insensitive (QDI) logic decomposition principles discussed in [39]. The resulting early output SAFA synthesized is depicted by Fig 4

$$\begin{aligned}SUM11 = &A11A01B10B00CIN0 + A10A01B11B00CIN0 + A11A00B10B01CIN0 + A10A00B11B01CIN0 \\ &+ A11A00B11B01CIN1 + A11A01B11B00CIN1 + A10A00B10B01CIN1 + A10A01B10B00CIN1 \\ &+ A10A01B10B01 + A11A00B10B00 + A10A00B11B00 + A11A01B11B01\end{aligned} \quad (5)$$

$$\begin{aligned}SUM10 = &A11A01B10B00CIN1 + A10A01B11B00CIN1 + A11A00B10B01CIN1 + A10A00B11B01CIN1 \\ &+ A10A01B10B00CIN0 + A10A00B10B01CIN0 + A11A01B11B00CIN0 + A11A00B11B01CIN0 \\ &+ A11A00B11B00 + A11A01B10B01 + A10A01B11B01 + A10A00B10B00\end{aligned} \quad (6)$$

$$SUM01 = A01B00CIN0 + A00B01CIN0 + A00B00CIN1 + A01B01CIN1 \quad (7)$$

$$SUM00 = A01B01CIN0 + A01B00CIN1 + A00B01CIN1 + A00B00CIN0 \quad (8)$$

$$\begin{aligned}COUT21 = &A10A00B11B01CIN1 + A11A00B10B01CIN1 + A10A01B11B00CIN1 + A11A01B10B00CIN1 \\ &+ A10A01B11B01 + A11A01B10B01 + A11B11\end{aligned} \quad (9)$$

$$\begin{aligned}COUT20 = &A11A01B10B00CIN0 + A10A01B11B00CIN0 + A11A00B10B01CIN0 + A10A00B11B01CIN0 \\ &+ A11A00B10B00 + A10A00B11B00 + A10B10\end{aligned} \quad (10)$$

Fig 6 Logic equations corresponding to the dual-rail encoded DAFA. The above equations are factorized [42] according to the safe QDI logic decomposition principles discussed in [39]. The resulting early output DAFA synthesized is portrayed by Fig 7

An early output SAFA based on the delay-insensitive dual-rail data encoding is shown in Fig 4 [11]. (A1, A0), (B1, B0), and (CIN1, CIN0) represent the dual-rail encoded augend, addend, and carry inputs. (SUM1, SUM0) and (COUT1, COUT0) are the dual-rail sum and carry outputs. The early output SAFA shown in Fig 4 consists of four complex gates (AO22) labelled as CG1 to CG4, four 2-input C-elements (highlighted by the circles with the marking C on their periphery), and two 2-input OR gates. The conjunction of A1 and B1 is visible in the complex gates CG1 and CG3 in Fig 4. Similarly, the conjunction of A0 and B0 is visible in the complex gates CG1 and CG4. This is an example of implicit logic redundancy [23] [38]. The logic equations corresponding to SAFA are expressed by (1) to (4) in Fig 5. Equations (1) to (4) are expressed in the disjoint sum-of-products (DSOP) form [34]. In a DSOP equation, the logical conjunction of any two products results in null [35] [36].

To realize a homogeneously encoded DAFA, the dual-rail code is used to encode the augend, addend, and carry inputs, and the sum and carry outputs. Let (A11, A10) and (A01, A00) represent the dual-rail augend inputs, and let (B11, B10) and (B01, B00) represent the dual-rail addend inputs, and let (CIN1, CIN0) represent the dual-rail carry input. The most significant and least significant dual-rail sum outputs are specified by (SUM11, SUM10) and (SUM01, SUM00) respectively. (COUT21, COUT20) denotes the dual-rail carry output. The logic equations of the DAFA outputs are given by (5) to (10), which are shown in Fig 6, and the gate level circuit is shown in Fig 7. Equations (5) to (10) also correspond to the DSOP form.

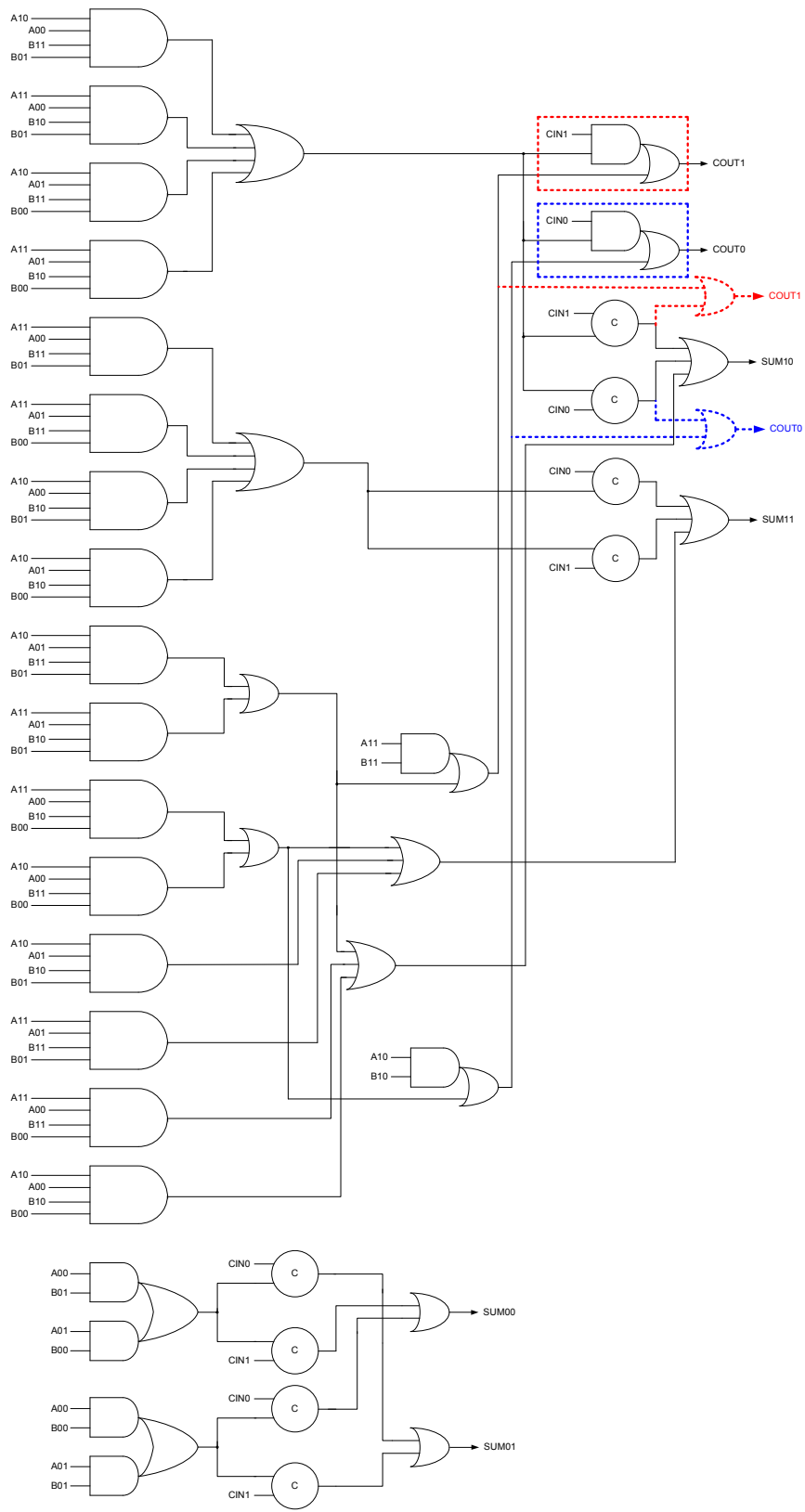

Fig 7 Early output DAFA [43] with or with no redundant logic with respect to the output carry

Fig 7 shows the early output DAFA circuit [43]. Fig 7 contains a mix of discrete gates, complex gates, and 2-input Muller C-elements. As described in [43], if the two complex gates viz. the AO21 gates shown within the red and blue rectangles in dotted lines in Fig 7 are removed, and if the two 2-input OR gates depicted in red and blue in dotted lines in Fig 7 are retained to synthesize COUT1 and COUT0 respectively, then the DAFA portrayed by Fig 7 would not have logic redundancy [38]. Alternatively, if the two 2-input OR gates shown in dotted lines in red and blue are removed, and if the two complex gates shown within the red and blue rectangles in dotted lines in Fig 7 are retained, then the DAFA would be said to contain redundant logic [38]. However, logic redundancy is implicit in the DAFA design as that of the SAFA design. Both the DAFA and SAFA designs would exhibit early reset behavior during the application of the spacer.

For the homogeneously encoded DAFA shown in Fig 7, which when positioned in an intermediate position in an RCA, the elements present in the critical path of the non-redundant design would be a 2-input C-element and a 2-input OR gate. On the other hand, the element present in the critical path of the redundant design would be just the AO21 gate. Hence the latency of the RCA embedding the DAFA with redundant logic would be less than the latency of the RCA containing the DAFA with no redundant logic. However, the introduction of logic redundancy may cause a slight increase in the area in the case of the former compared to the latter.

IV. PHYSICAL IMPLEMENTATION AND DESIGN PARAMETERS OF DIFFERENT ASYNCHRONOUS ADDERS AND COMPARISON OF THEIR THEORETICAL AND PRACTICAL LATENCIES

Many 32-bit asynchronous adders such as RCAs [10] [11] [17] [18] [38] [43], CLAs [44] [45], and an optimized CSLA [41] were implemented for comparison based on a 32/28nm CMOS process [37]. The asynchronous RCAs were physically realized by cascading homogeneously or heterogeneously encoded SAFAs and/or DAFAs corresponding to weak-indication and early output types. The circuit architectures of the homogeneously and heterogeneously encoded asynchronous adders are given in [38], and the interested reader is referred to the same for details. The 2-input Muller C-element was manually implemented using 12 transistors and it was made available to physically implement the various asynchronous adders, registers, and the completion detector. High fan-in C-element functionality wherever imminent was safely decomposed into a 2-input C-element tree based on the QDI logic decomposition method presented in [39] which guarantees gate-orphan freedom.

An asynchronous circuit stage, as shown in Fig 1, consists of the asynchronous function block, the input registers, and the completion detector. With respect to the asynchronous function blocks realized on the basis of heterogeneous delay-insensitive data encoding, dual-rail to 1-of-4 encoders would be present before the function block and 1-of-4 to dual-rail decoders would be present after the function block as illustrated in [38]. The input registers and the completion detector would be identical and only the asynchronous function blocks would tend to differ in their physical composition. Hence any differences between the simulation results of the various asynchronous adders can be attributed to the differences between their respective function blocks. This paves the way for a straightforward comparison of the design metrics viz. latency, area, and average power dissipation of the different asynchronous adders after physical synthesis.

About 1000 random input vectors were identically supplied to all the asynchronous adders at time intervals of 20ns through test benches to verify their functionalities and also to capture their respective switching activities. The .vcd files generated through the functional simulations were subsequently used for average power estimation using Synopsys PrimeTime. Since the EDA tool, by default, estimates the critical path timing, the worst-case forward latency was alone estimated for a typical case PVT specification (1.05V, 25ºC) of the standard cell library [37]. Appropriate wire load was automatically inserted into the adders while performing the simulations. A virtual clock was used to just constrain the input and output ports of the asynchronous adders and it did not contribute to the circuit power dissipation. Table 2 presents the simulation results viz. critical path delay (i.e. forward latency), area occupancy, and average power dissipation of various 32-bit asynchronous adders, and the optimum design metrics are highlighted in bold face. Note the use of adder legends viz. Adder1 to Adder18 in Table 2 to simplify the discussion of results.

Two general observations can be made from Table 2. Homogeneously encoded asynchronous adders feature reduced latency, area, and power dissipation than the heterogeneously encoded asynchronous adders. It was shown in [43] that the homogeneously encoded asynchronous RCAs incorporating DAFAs report optimized design metrics than their heterogeneously encoded RCA counterparts containing DAFAs. While this may be due to the difference in the function blocks, the presence of extra encoders and decoders in the case of the heterogeneously encoded asynchronous circuit causes a disadvantage compared to the homogeneously encoded circuit counterpart. The proposed asynchronous RCA, referred to as Adder11 in Table 2, containing 15 DAFAs and 2 SAFAs reports the least latency among all the 32-bit asynchronous adders. The introduction of extra SAFAs into the least significant positions as a replacement for the DAFAs tends to increase the latency although this may reduce the area, as noticed in the case of Adder12 in Table 2. Hence, SAFAs cannot be arbitrarily used to replace the DAFAs to minimize the latency, and the replacement should be guided say on the basis of a static timing analysis.

A theoretical computation of the (forward) latency of different 32-bit asynchronous adders mentioned in Table 2 is performed and is correlated with the practical latency estimates. Equations (11) to (27) given in Fig 8 represent the approximate theoretical expressions governing the latency of various 32-bit asynchronous adders considered. The equations are termed approximate since they capture only the typical gate delays and not the wire delays. However the practical latency estimates given in Table 2 include both gate delays and wire delays. An additional buffer delay of 0.04ns was included while computing the theoretical latency estimates to maintain parity with the practical latency estimates.

TABLE 2 DESIGN METRICS OF DIFFERENT 32-BIT ASYNCHRONOUS ADDERS, ESTIMATED USING A 32/28NM CMOS PROCESS. THE ADDERS CORRESPOND TO WEAK-INDICATION OR EARLY OUTPUT TYPES

| Adder legend | Adder type; Encoding; Timing model | Latency (ns) | Area ($\mu m^2$) | Power ($\mu W$) |
|---|---|---|---|---|
| Adder1 | RCA [11]; Homogeneous, Redundant logic; Early output | 3.10 | **1658.80** | **2161** |
| Adder2 | RCA [10]; Heterogeneous, No redundancy; Weak-indication | 7.06 | 2016.63 | 2170 |
| Adder3 | RCA [17, 38]; Homogeneous, No redundancy; Weak-indication | 4.12 | 2866.49 | 2200 |
| Adder4 | RCA [17, 38]; Homogeneous, Redundant logic; Weak-indication | 2.84 | 2931.55 | 2202 |
| Adder5 | RCA [43]; Homogeneous, No redundancy; Early output | 4.01 | 2472.06 | 2174 |
| Adder6 | RCA [43]; Homogeneous, Redundant logic; Early output | 2.21 | 2488.32 | 2173 |
| Adder7 | RCA [18, 38]; Heterogeneous, No redundancy; Weak-indication | 4.36 | 3301.58 | 2191 |
| Adder8 | RCA [18, 38]; Heterogeneous, Redundant logic; Weak-indication | 3.03 | 3366.65 | 2192 |
| Adder9 | RCA [43]; Heterogeneous, No redundancy; Early output | 4.22 | 2634.71 | 2182 |
| Adder10 | RCA [43]; Heterogeneous, Redundant logic; Early output | 2.38 | 2650.98 | 2182 |
| Adder11 | Proposed RCA; Homogeneous, Redundant logic; 15 DAFAs and 2 SAFAs; Early output | **2.14** | 2436.48 | 2173 |
| Adder12 | Proposed RCA; Homogeneous, Redundant logic; 14 DAFAs and 4 SAFAs; Early output | 2.21 | 2384.63 | 2171 |
| Adder13 | CLA [44]; Homogeneous; Weak-indication | 3.31 | 2951.88 | 2191 |
| Adder14 | CLA-RCA [44]; Homogeneous; Weak-indication | 3.08 | 2845.14 | 2189 |
| Adder15 | CLA [45]; Homogeneous; Early output | 2.77 | 2569.65 | 2177 |
| Adder16 | CLA-RCA [45]; Homogeneous; Early output | 2.54 | 2455.80 | 2175 |
| Adder17 | CSLA [41]; Homogeneous; Early output | 2.46 | 3000.17 | 2293 |

The theoretical and practical latency estimates are normalized based on the minimum latency value reported for Adder11, which is the proposed 32-bit asynchronous RCA that corresponds to the early output type. The normalized values of theoretical and practical latencies corresponding to the various 32-bit asynchronous adders mentioned in Table 2 are portrayed side-by-side in Fig 9 for comparison. A 2-point moving average trend line is plotted in dashed lines in Fig 9 with respect to the theoretical and practical latencies to aid the comparison. It is clear that there is a close correlation between the normalized magnitudes of the theoretical and practical latencies corresponding to the various asynchronous adders which vindicates the equations presented in Fig 8.

It is noted from Table 2 that with respect to area and power dissipation, Adder1 is optimized than the rest. This is quite expected since the SAFA [11] constituting Adder 1 requires just 27.45$\mu m^2$ of silicon. On the other hand, the DAFA [43] requires 106.74$\mu m^2$ of silicon, which signifies almost a 3× area increase compared to the area occupancy of the SAFA. The less area occupancy of the SAFA shown in Fig 4 compared to the DAFA shown in Fig 7 translates into a marginally less average power dissipation for Adder1 compared to Adder11 (i.e. 0.5%). The reduction in power dissipation is only meagre though and this is because all the asynchronous adders mentioned in Table 2 tend to satisfy the monotonic cover constraint [19], which implies the activation of a unique signal path from a primary input to a primary output. The monotonic cover constraint results from the DSOP expressions of the adder outputs. However in terms of latency, Adder11 is optimized than Adder1 by 31%. It is noted that the proposed early output 32-bit asynchronous RCA, i.e. Adder11 which comprises 2 stages of SAFAs in the least significant positions and 15 stages of DAFAs in the more significant positions achieves the following optimizations in latency over its architectural counterparts for a similar adder size: (i) 35.3% reduction in latency over a weak-indication SCBCLA [44], (ii) 30.5% reduction in latency over a weak-indication hybrid SCBCLA-RCA [44], (iii) 20.2% reduction in latency over an early output RCLA [45], (iv) 18.7% reduction in latency over an early output hybrid RCLA-RCA [45], and (v) a 13% reduction in latency over an early output CSLA that features an optimum 8-8-8-8 uniform input partition [41].

$$T_{Adder1} = T_{BUF} + T_{REG} + 32T_{AO22} + T_{CE2} + T_{OR2} \quad (11)$$

$$T_{Adder2} = T_{BUF} + T_{REG} + 32T_{CE2} + 33T_{OR2} \quad (12)$$

$$T_{Adder3} = T_{BUF} + T_{REG} + 16T_{CE2} + T_{AND4} + T_{OR4} + T_{OR3} + 15T_{OR2} \quad (13)$$

$$T_{Adder4} = T_{BUF} + T_{REG} + T_{CE2} + T_{AND4} + 15T_{AND2} + T_{OR4} + T_{OR3} + 15T_{OR2} \quad (14)$$

$$T_{Adder5} = T_{BUF} + T_{REG} + 16T_{CE2} + T_{AND4} + T_{OR4} + T_{OR3} + 15T_{OR2} \quad (15)$$

$$T_{Adder6} = T_{BUF} + T_{REG} + 15T_{AO21} + T_{CE2} + T_{AND4} + T_{OR4} + T_{OR3} \quad (16)$$

$$T_{Adder7} = T_{BUF} + T_{REG} + 17T_{CE2} + 18T_{OR2} \quad (17)$$

$$T_{Adder8} = T_{BUF} + T_{REG} + 2T_{CE2} + 15T_{AND2} + 18T_{OR2} \quad (18)$$

$$T_{Adder9} = T_{BUF} + T_{REG} + T_{AO22} + 16T_{CE2} + 17T_{OR2} \quad (19)$$

$$T_{Adder10} = T_{BUF} + T_{REG} + 15T_{AO21} + T_{CE2} + T_{AND2} + T_{OR4} + T_{OR2} \quad (20)$$

$$T_{Adder11} = T_{BUF} + T_{REG} + 3T_{AO22} + 14T_{AO21} + T_{CE2} + T_{OR3} \quad (21)$$

$$T_{Adder12} = T_{BUF} + T_{REG} + 5T_{AO22} + 13T_{AO21} + T_{CE2} + T_{OR3} \quad (22)$$

$$T_{Adder13} = T_{BUF} + T_{REG} + 12T_{CE2} + 3T_{AO222} + T_{AND4} + 2T_{OR4} + 8T_{OR2} \quad (23)$$

$$T_{Adder14} = T_{BUF} + T_{REG} + 11T_{CE2} + 3T_{AO222} + T_{AND4} + 2T_{OR4} + 7T_{OR2} \quad (24)$$

$$T_{Adder15} = T_{BUF} + T_{REG} + 12T_{CE2} + T_{AO22} + 9T_{OR2} \quad (25)$$

$$T_{Adder16} = T_{BUF} + T_{REG} + 11T_{CE2} + T_{AO22} + 8T_{OR2} \quad (26)$$

$$T_{Adder17} = T_{BUF} + T_{REG} + 6T_{CE2} + 9T_{AO22} + 3T_{OR2} \quad (27)$$

Fig 8 Simplified theoretical expressions governing the latency of different 32-bit asynchronous adders mentioned in Table 2

Note: In the above equations, the notation T represents the typical propagation delay. The individual gate delays are typical delay values corresponding to the minimum size gates present in the cell library [37].

$T_{AdderX}$ refers to the typical datapath delay, i.e. the forward latency of a generic asynchronous adder with the legend 'AdderX'. $T_{BUF}$ is the delay of the non-inverting buffer. $T_{REG}$ refers to the delay of a register element, which is equal to the delay of a 2-input C-element ($T_{CE2}$). $T_{AND2}$ is the delay of a 2-input AND gate. $T_{AND4}$ is the delay of a 4-input AND gate. $T_{OR2}$ is the delay of a 2-input OR gate. $T_{OR3}$ is the delay of a 3-input OR gate. $T_{OR4}$ is the delay of a 4-input OR gate. $T_{AO222}$ is the delay of the AO222 complex gate. $T_{AO22}$ is the delay of the AO22 complex gate, and $T_{AO21}$ is the delay of the AO21 complex gate.

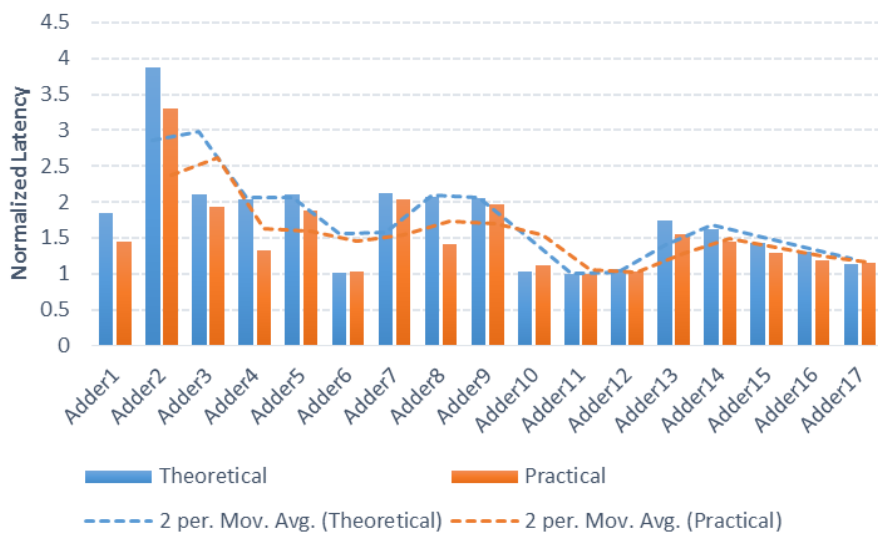

Fig 9 Comparison of normalized theoretical and practical latency estimates of diverse 32-bit asynchronous adders. The adder legends mentioned in Table 2 are specified in the X-axis

V. CONCLUSION

This paper has presented a new asynchronous early output RCA (and RCA architecture) that incorporates DAFAs and SAFAs to achieve significant optimization in the latency. The DAFAs and SAFAs incorporate redundant logic and correspond to the early output type and are based on the homogeneous delay-insensitive dual-rail data encoding scheme which obey a 4-phase handshaking. The optimum number of SAFAs and DAFAs to be used for the least significant and more significant adder positions would in fact depend on the adder size. In this work, a 32-bit dual-operand addition was considered and it was found that employing 2 stages of SAFAs in the least significant positions and 15 stages of DAFAs in the more significant positions leads to an optimized latency over the other architectural counterparts such as CLA and CSLA.

A number of 32-bit asynchronous adders were constructed based on RCA, CLA, and CSLA architectures and their design metrics viz. latency, area, and average power dissipation were estimated. Also, the theoretical latencies of the asynchronous adders were computed based on a mathematical modeling of the critical path delay, and the normalized theoretical and practical latency values were compared which showed a close correlation.

In general, it appears that homogeneous delay-insensitive data encoding is preferable than heterogeneous delay-insensitive data encoding for a robust asynchronous circuit design. Adder1, mentioned in Table 2, reports the least area and power dissipation while Adder 11 reports the optimized latency, and both Adder1 and Adder11 employ homogeneous delay-insensitive data encoding based on the dual-rail code. While this phenomenon may be partly due to the differences in the function block composition based on homogeneous and heterogeneous delay-insensitive data codes, the other reason could be that the latter requires the provision of extra protocol conversion circuits, i.e. the dual-rail to 1-of-4 encoder and the 1-of-4 to dual-rail decoder, before and after an asynchronous circuit stage, and this tends to deteriorate the circuit design metrics. In the future, the utility of the proposed asynchronous RCA (or the RCA architecture) for the effective realization of multi-operand additions [46] could be considered since multi-operand additions are predominant in digital signal processing applications.